# Competition between singlet and triplet superconductivity

Tian De Cao, Tie Bang Wang

*Department of Physics, Nanjing University of Information Science & Technology, Nanjing 210044, China*

**Abstract**

The competition between singlet and triplet superconductivity is examined in consideration of correlations on an extended Hubbard model. It is shown that the triplet superconductivity may not be included in the common Hubbard model since the strong correlation favors the singlet superconductivity, and thus the triplet superconductivity should be induced by the electron-phonon interaction and the ferromagnetic exchange interaction. We also present a superconducting qualification with which magnetism is unbeneficial to superconductivity.

## 1. Introduction

Some experiments argue that superconductivity could appear in the magnetism ordered phase [1-3], and theoretic works also suggest that superconductivity might coexist with magnetism [4-8], but there have been other suggestions [9-10]. One finds that the high temperature superconductivity usually appears in the border of the magnetic orders [11], and these examples include Cu-based superconductors [12-15] and Fe-based superconductors [16-20]. Therefore, the relation between superconductivity and magnetism, and the competition between singlet and triplet superconductivity, should be an interesting topic.

## 2. BCS approximation

To consider the competition between singlet and triplet superconductivity, we are interested in the extended

Hubbard model

$$H = \sum_{l,l',\sigma}(t_{ll'} - \mu\delta_{ll'})d^+_{l\sigma}d_{l'\sigma} + U\sum_{l} n_{l\uparrow}n_{l\downarrow} + \frac{1}{4}\sum_{l,l',\sigma,\sigma'}V_{ll'}n_{l\sigma}n_{l'\sigma'} - \sum_{l,l'}J_{ll'}\hat{S}_{lz}\hat{S}_{l'z} \quad (1)$$

After introducing the charge operator $\hat{\rho}(q) = \frac{1}{2}\sum_{k,\sigma}d^+_{k+q\sigma}d_{k\sigma}$ and the spin operator $\hat{S}(q) = \frac{1}{2}\sum_{k,\sigma}\sigma d^+_{k+q\sigma}d_{k\sigma}$ in the wave vector space, we write the model (1) in

$$H = \sum_{k,\sigma}\xi_k d^+_{k\sigma}d_{k\sigma} + \sum_{q} V(q)\hat{\rho}(q)\hat{\rho}(-q) - \sum_{q} J(q)\hat{S}_z(q)\hat{S}_z(-q) \quad (2)$$

where $\xi_k = \varepsilon_k - \mu$, $V(q) = U + V_0(q)$, $J(q) = U + J_0(q)$, and we denote wave vector $\vec{k}$ as $k$, $k \equiv \vec{k}$. Equation.(2) shows that the on-site interaction U also contributes a ferromagnetic coupling. The symmetry of crystal lattice leads to $V(\bar{q}) = V(q)$, $J(\bar{q}) = J(q)$ and $\xi_{\bar{k}} = \xi_k$, thus the wave vector dependence of each physical quantity meets $f(\bar{k}) = f(k)$. The effects of phonons are assumed to be included in $V(q)$.

Define these Green's functions

$$G(k\sigma, \tau - \tau') = -<T_\tau d_{k\sigma}(\tau)d^+_{k\sigma}(\tau')> \quad (3)$$

$$F^+(k\sigma, \tau - \tau') = <T_\tau d^+_{k\sigma}(\tau)d^+_{\bar{k}\sigma}(\tau')> \quad (4)$$

$$F(k\sigma, \tau - \tau') = <T_\tau d_{\bar{k}\sigma}(\tau)d_{k\sigma}(\tau')> \quad (5)$$

Without considering the effect of strong correlation, we arrive at the BCS-like equation

$$\Delta(k\sigma) = \sum_{q}[V(k-q) - J(k-q)]\Delta(q\sigma) \cdot \frac{n_F(E_{q\sigma}) - n_F(-E_{q\sigma})}{2E_{q\sigma}} \quad (6)$$

$$G(k\sigma, \tau = 0) = \frac{E_{k\sigma} - \tilde{\xi}_{k\sigma}}{2E_{k\sigma}}n_F(-E_{k\sigma}) + \frac{E_{k\sigma} + \tilde{\xi}_{k\sigma}}{2E_{k\sigma}}n_F(E_{k\sigma}) \quad (7)$$

where

$$\tilde{\xi}_{k\sigma} = \xi_k - \sigma s_{lz}J(0) - \frac{1}{2}\sum_{q}[V(q) - J(q)]n_{k+q,\sigma} \quad (8)$$

$$\Delta(k\sigma) = -\frac{1}{2}\sum_{q}[V(q) - J(q)]F^+(k-q, \sigma, \tau = 0) \quad (9)$$

and $E_{q\sigma} = \sqrt{\tilde{\xi}^2_{q\sigma} + \Delta^2(q\sigma)}$.

If we consider the singlet superconductivity, the functions (4) and (5) have to be redefined. For example, we should define $F^+(k\sigma, \tau - \tau') = <T_\tau d_{k\sigma}^+(\tau) d_{\bar{k}\bar{\sigma}}^+(\tau')>$, and the $V(q) - J(q)$ in Eqs.(6) and (9) should be changed into $V(q) + J(q)$. Equation (9) and its application on the singlet superconductivity arrive at these possible conclusions (without effect of strong correlation):

(Ⅰ) If $V_{ll'}=0$ and $J_{ll'}=0$, the triplet superconductivity could not occur.

(Ⅱ) Whether the singlet or triplet superconductivity is favored could not be discriminated if $J=0$ and $U=0$.

(Ⅲ) The singlet superconductivity is forbidden for a large $U$, while the triplet superconductivity is allowed.

(Ⅳ) Ferromagnetic exchange parameter $J$ favors the triplet superconductivity, while antiferromagnetic exchange parameter $J$ favors the singlet superconductivity.

## 3. Effects of strong correlation

To consider the effects of correlations, we must calculate many-particle correlation functions such as $\partial_\tau <T_\tau \hat{S}(q) d_{k+q\sigma} d_{k\sigma}^+(\tau')>$ and $\partial_\tau <T_\tau \hat{\rho}(q) d_{k+q\sigma} d_{k\sigma}^+(\tau')>$. In higher level of approximation we obtain the equations

$$[-i\omega_n + \tilde{\xi}_{k\sigma} + \sum_q \frac{P(k,q,\sigma)}{i\omega_n - \xi_{k+q}}] G(k\sigma, i\omega_n)$$

$$= -1 + \frac{I_\sigma}{-i\omega_n + \xi_k} + \frac{1}{2}\sum_q \frac{\xi_{k+q} - \xi_k}{-i\omega_n + \xi_{k+q}} [V_0(q) - J_0(q)] F(k+q\sigma, \tau=0) F^+(k\sigma, i\omega_n) \quad (10)$$

and

$$[-i\omega_n - \tilde{\xi}_{k\sigma} - \sum_q \frac{P(k,q,\sigma)}{i\omega_n + \xi_{k+q}}] F^+(k\sigma, i\omega_n)$$

$$= \frac{1}{2}\sum_q \frac{\xi_{k+q} - \xi_k}{-i\omega_n - \xi_{k+q}} [V_0(q) - J_0(q)] F^+(k-q, \sigma, \tau=0) G(k\sigma, i\omega_n) \quad (11)$$

where $I_\sigma = \sigma J(0) s_{lz} + V(0) <\hat{\rho}(0)>$, and

$$P(k,q,\sigma) = \frac{1}{2}(\xi_{k+q} - \xi_k)(-J(q) + V(q)) n_{k+q,\sigma} + J(-q) <\hat{S}(-q)\hat{S}(q)> J(q)$$

$$-2\sigma V(-q)<\hat{\rho}(-q)\hat{S}(q)>J(q)+V(-q)<\hat{\rho}(-q)\hat{\rho}(q)>V(q) \tag{12}$$

The function $P(k,q,\sigma)$ exhibits the effects of correlations. $<\hat{S}(-q)\hat{S}(q)>$ is the spin-spin correlation function at equal time; other correlation functions are similar to this one. It is found that both the on-site interaction $U$ and the ferromagnetic coupling $J_{ll'}$ will increase the spin dependence of $P(k,q,\sigma)$ in Eq.(12), they strengthen the spin correlation, thus a solution of ferromagnetism is easy to obtain with Eq.(10). At the same time, it is also shown that $V_{ll'}$ is beneficial for the charge correlation. For simplification, we consider $T<T_c$ and $T\to T_c$ and get

$$F^+(k\sigma,\tau=0)=-\frac{1}{\beta}\sum_n [i\omega_n+\tilde{\xi}_{k\sigma}+\Sigma^{(+)}(k,i\omega_n)]^{-1}\cdot\frac{\Delta^{(-)}_+(k,i\omega_n)}{i\omega_n-\tilde{\xi}_{k\bar{\sigma}}-\Sigma^{(-)}(k,i\omega_n)}(1-\frac{I_\sigma}{-i\omega_n+\xi_k}) \tag{13}$$

where

$$\Sigma^{(\pm)}(k,i\omega_n)=\sum_q \frac{P(k,q,\sigma)}{i\omega_n\pm\xi_{k+q}}$$

$$\Delta^{(\pm)}(k,i\omega_n)=\frac{1}{2}\sum_q \frac{\xi_{k+q}-\xi_k}{-i\omega_n\pm\xi_{k+q}}[V(q)-J(q)]F(k+q\sigma,\tau=0) \tag{14}$$

$$\Delta^{(\pm)}_+(k,i\omega_n)=\frac{1}{2}\sum_q \frac{\xi_{k+q}-\xi_k}{-i\omega_n\pm\xi_{k+q}}[V(q)-J(q)]F^+(k+q\sigma,\tau=0)$$

To obtain an evident solution, we assume the on-site interaction $U$ is not too large. Therefore, the function $F^+$ dominated by the frequency region where $\text{Im}\Sigma^{(+)}(k,\omega)=0$ is determined by

$$F^+(k\sigma,\tau=0)=\frac{1}{2}\frac{1}{E_{k\sigma,1}-E_{k\sigma,2}}[n_F(E_{k\sigma,1})\ z^{(+)}(E_{k\sigma,1})\ (1+\frac{I_\sigma}{E_{k\sigma,1}-\xi_k})\sum_q \frac{\xi_{k+q}-\xi_k}{\xi_{k+q}+E_{k\sigma,1}}$$

$$-n_F(E_{k\sigma,2})\ z^{(-)}(E_{k\sigma,2})\ (1+\frac{I_\sigma}{E_{k\sigma,2}-\xi_k})\sum_q \frac{\xi_{k+q}-\xi_k}{\xi_{k+q}+E_{k\sigma,2}}][V(q)-J(q)]F^+(k+q\sigma,\tau=0) \tag{15}$$

where the spectral weight $z^{(\pm)}(\omega)=[1+\sum_q \frac{P(k,q,\sigma)}{(\omega\pm\xi_{k+q})^2}]^{-1}$. $\omega=E_{k\sigma,1}$ expresses the real solution of $\omega+\tilde{\xi}_{k\sigma}+\text{Re}\Sigma^{(+)}(k,\omega)=0$, and $\omega=E_{k\bar{\sigma},2}$ expresses the real solution of $\omega-\tilde{\xi}_{k\bar{\sigma}}-\text{Re}\Sigma^{(-)}(k,\omega)=0$, in which the spin index dependences are duo to the spin-charge correlation in Eq. (12).

It is hard to obtain the obvious forms of both $E_{k\sigma,1}$ and $E_{k\bar{\sigma},2}$, while we can find them by the successive iteration method. For example, $E_{k\sigma,1}^{(0)} = -\tilde{\xi}_{k\sigma}$, $E_{k\sigma,1}^{(1)} = -\tilde{\xi}_{k\sigma} + \sum_q \frac{P(k,q,\sigma)}{\tilde{\xi}_{k\sigma} - \xi_{k+q}}$, and so on. There are superconducting solutions for $F^+ \neq 0$ when $T_c > 0$K for either negative or positive $V(q) \pm J(q)$. This can be easily found at the Fermi surface where we have $E_{k_F\sigma,1} = E_{k_F\bar{\sigma},2} \to 0 (k_F \equiv \vec{k}_F)$. Similarly, the $V(q) - J(q)$ in Eq.(15) should be changed into $V(q) + J(q)$ for the singlet pairing. After a careful analysis, we arrive at these possible conclusions:

(1) If $V_{ll'} = 0$ and $J_{ll'} = 0$, the triplet superconductivity could not occur.

(2) Whether the singlet or triplet superconductivity is favored could not be discriminated if $J = 0$ and $U = 0$.

(3) The singlet superconductivity is favored for a large $U$, while the triplet superconductivity is also allowed.

(4) Both ferromagnetic and antiferromagnetic $J$ favor the singlet and triplet superconductivity.

## 4. Discussions and conclusions

Comparing the conclusions(Ⅰ)-(Ⅳ) with the conclusions (1)-(4), we find the conclusion (Ⅰ) is just the conclusion(1), and the conclusion (Ⅱ) is just the conclusion (2). That is to say, the conclusions (Ⅰ) and (Ⅱ) are not changed with or without the effect of strong correlation. The conclusion (Ⅱ) is in fact the result under weak correlation, and it is shown that whether the singlet or triplet superconductivity is favored could not be discriminated when the effect of correlation is neglected. The conclusion (Ⅰ) can be re-expressed: The triplet superconductivity could not be included in the (comm. on) Hubbard model. The conclusions (Ⅲ)-(Ⅳ) and (3)-(4) should be corrected under considering the results in experiments. It is shown that the singlet superconductivity is favored for a large $U$, thus the triplet superconductivity is not from the effect of strong correlation and this arrives at this conclusion (Ⅳ). Now let us rewrite the results as (a)-(d) below:

(a) The triplet superconductivity could not be included in the (common) Hubbard model.

(b) Whether the singlet or triplet superconductivity is favored could not be discriminated without considering

the effects of strong correlations.

(c) The strong correlation favors the singlet superconductivity.

(d) The triplet superconductivity should originate from the effect of both the electron-phonon interaction and the ferromagnetic exchange interaction.

We now discuss the effect of possible magnetism on superconductivity in a simple way. When magnetisms exist in the superconductors, the spin-charge correlation term in Eq. (12) must be not zero. The singlet superconductivity meets $|\Delta(k\sigma)| = |\Delta(k\bar{\sigma})|$ because $|<|d_\uparrow d_\downarrow|>| = |<|d_\downarrow d_\uparrow|>|$ for any state $|>$. The triplet superconductivity should also meet $|\Delta(k\sigma)| = |\Delta(k\bar{\sigma})|$ since the gaps of $|\Delta(k\sigma)| \neq |\Delta(k\bar{\sigma})|$ have not been observed in triplet superconductors. Because the temperature determined by both $|\Delta(k\sigma)| = |\Delta(k\bar{\sigma})|$ and gap equation is lower than the one determined by the gap equation, it is easy to understand that the extra condition $|\Delta(k\sigma)| = |\Delta(k\bar{\sigma})|$ usually decreases $T_c$ when magnetism exists in superconductors, our theoretic prediction in superconductivity must follow this qualification.

In summary, we have arrived at the conclusions (a)-(d) and suggested the qualification $|\Delta(k\sigma)| = |\Delta(k\bar{\sigma})|$, and the qualification shows that magnetism intends to weaken the superconductivity.

**ACKNOWLEDGMENTS**

We thank Nanjing University of Information Science & Technology for financial support.